\documentstyle[11pt,aaspp4,flushrt]{article}
\begin{document}
\newcommand{\lya}{Lyman~$\alpha$}
\newcommand{\lyb}{Lyman~$\beta$}
\newcommand{\za}{$z_{\rm abs}$}
\newcommand{\ze}{$z_{\rm em}$}
\newcommand{\cmtwo}{cm$^{-2}$}
\newcommand{\nhi}{$N$(H$^0$)}
\newcommand{\nzn}{$N$(Zn$^+$)}
\newcommand{\ncr}{$N$(Cr$^+$)}
\newcommand{\degpoint}{\mbox{$^\circ\mskip-7.0mu.\,$}}
\newcommand{\halpha}{\mbox{H$\alpha$}}
\newcommand{\hbeta}{\mbox{H$\beta$}}
\newcommand{\hgamma}{\mbox{H$\gamma$}}
\newcommand{\kms}{\,km~s$^{-1}$}      
\newcommand{\minpoint}{\mbox{$'\mskip-4.7mu.\mskip0.8mu$}}
\newcommand{\mv}{\mbox{$m_{_V}$}}
\newcommand{\Mv}{\mbox{$M_{_V}$}}
\newcommand{\peryr}{\mbox{$\>\rm yr^{-1}$}}
\newcommand{\secpoint}{\mbox{$''\mskip-7.6mu.\,$}}
\newcommand{\sqdeg}{\mbox{${\rm deg}^2$}}
\newcommand{\squig}{\sim\!\!}
\newcommand{\subsun}{\mbox{$_{\twelvesy\odot}$}}
\newcommand{\et}{et al.~}

\def\ltsima{$\; \buildrel < \over \sim \;$}
\def\simlt{\lower.5ex\hbox{\ltsima}}
\def\gtsima{$\; \buildrel > \over \sim \;$}
\def\simgt{\lower.5ex\hbox{\gtsima}}
\def\arcs{$''~$}
\def\arcm{$'~$}
\title{SPECTROSCOPIC CONFIRMATION OF A POPULATION OF NORMAL
STAR FORMING GALAXIES AT REDSHIFTS Z${\bf >}$3\altaffilmark{1}}
\author{\sc Charles C. Steidel\altaffilmark{2,3,4}, Mauro Giavalisco\altaffilmark{5,6},
Max Pettini\altaffilmark{7}, Mark Dickinson\altaffilmark{8}, and Kurt L.
Adelberger\altaffilmark{2}}

\altaffiltext{1}{Based in large part on observations obtained at the
W.M. Keck Observatory, which is operated jointly by the California Institute
of Technology and the University of California }
\altaffiltext{2}{Palomar Observatory, California Institute of Technology,
Mail Stop 105-24, Pasadena, CA 91125}
\altaffiltext{3}{Alfred P. Sloan Foundation Fellow}
\altaffiltext{4}{NSF Young Investigator}
\altaffiltext{5}{Observatories of the Carnegie Institution of Washington, 813
Santa Barbara Street, Pasadena, CA 91101}
\altaffiltext{6}{Hubble Fellow}
\altaffiltext{7}{Royal Greenwich Observatory, Madingley Road, Cambridge CB3 OEZ, UK}
\altaffiltext{8}{Space Telescope Science Institute, 3700 San Martin Drive,
Baltimore, MD 21218}

\begin{abstract}

We report the discovery of a substantial population of 
star--forming galaxies at $3.0 \simlt z \simlt 3.5$.
These galaxies have been selected using color criteria sensitive to the 
presence of a Lyman continuum break superposed on an otherwise very blue 
far-UV continuum, and then confirmed with deep spectroscopy on
the W. M. Keck telescope. 
The surface density of galaxies brighter than ${\cal R} = 25$
with $3.0 \simlt z \simlt 3.5$ is $0.4\pm0.07$ galaxies arcmin$^{-2}$, 
approximately 1.3\% of the deep counts at these magnitudes; this value 
applies both to ``random'' fields and to fields centered 
on known QSOs.
The corresponding 
co-moving space density 
is approximately half that of luminous ($L \simgt L^{\ast}$) present--day galaxies. 
Our sample of $z > 3$ galaxies is large enough that we 
can begin to detail the spectroscopic characteristics of the population 
as a whole. The spectra of 
the $z>3$ galaxies  
are remarkably similar to those of nearby star--forming galaxies,
the dominant features being strong low--ionization interstellar 
absorption lines and high--ionization stellar lines, often with P-Cygni 
profiles characteristic of Wolf-Rayet and O--star winds.
Lyman $\alpha$
emission is generally weak ($< 20$ \AA\ rest equivalent width)
and is absent for $>$50\% of the galaxies.
We assign
approximate mass scales to the galaxies using the strengths of
the heavily saturated interstellar features and find
that, if the line widths are dominated by  
gravitational motions within the galaxies, 
the implied velocity dispersions are
$180 \le \sigma \le 320$ \kms, in the range expected
for massive galaxies.
The star formation rates, which can be
measured directly from the far-UV continua, lie in the range
$4 - 25$ $h_{50}^{-2}$ $M_{\sun}$ yr$^{-1}$ (for
$q_0=0.5$), with $8.5h_{50}^{-2}$ $M_{\sun}$ yr$^{-1}$ being typical. 
Together with the morphological properties
of the $z>3$ galaxy population, which we discuss in a companion paper
(Giavalisco \et 1996), all of these findings strongly suggest that we
have identified the high-redshift counterparts of the 
spheroid component of present--day 
luminous galaxies. In any case, it is clear that massive galaxy formation
was already well underway by $z \sim 3.5$.  
\end{abstract}

\section{INTRODUCTION}

Several years ago we embarked on a program to assess the presence of ``normal''
star--forming galaxies at redshifts $z >3$ using a custom suite of 
broad--band filters (dubbed $U_n$, $G$, and ${\cal R}$-- see Steidel \& Hamilton
1993) designed to reach 
very faint apparent magnitudes with accurate colors.
The technique rests on only two assumptions about the spectra of 
high-redshift galaxies: {\it (i)} a roughly flat  
spectrum ($f_{\nu} \propto \nu^0$) in the far-UV 
where the flux is dominated by emission from massive stars and, 
{\it (ii)} a pronounced spectral ``break'' at
912 \AA\ (the Lyman limit in the galaxy rest frame)
due to the combination of the intrinsic
stellar energy distributions, the inherent opacity of the galaxies
to Lyman continuum photons, and the effects of intervening
absorbers
(see Steidel \& Hamilton 1992, 1993; Steidel, Pettini,
\& Hamilton 1995 [Papers I,II,III] for an extensive
discussion of the technique; see also Madau 1995). 
The filters are ``tuned'' to selecting
objects with redshifts in the range $3.0 \simlt z \simlt 3.5$, where
the $U_n$ passband falls mostly shortward of the Lyman limit and
the $G$ band is not severely blanketed by the Lyman $\alpha$
forest---at higher redshifts this blanketing can make it very difficult
to distinguish a spectral break from
an intrinsically red spectral energy distribution.

While earlier work 
(Guhathakurta \et 1990) had shown that Lyman break galaxies
do not {\it dominate} the galaxy counts at faint apparent magnitudes, 
these constraints were too weak to rule out the presence of a 
substantial population of such objects among
the more prominent foreground blue galaxies.
Our first searches were directed at the fields of QSOs
with known Lyman limit absorption systems at $z > 3$;
these  could then be used
as ``templates'' to guide one (eventually) to galaxies in the general field, 
or at least to test the feasibility of detecting normal galaxies
at such extreme redshifts. 
In Paper III we discussed together the properties of both QSO absorbers
and candidates in the general field, since QSO absorbers are apparently 
drawn from the population of relatively luminous (but otherwise normal) field 
galaxies, at least at smaller redshifts (see Steidel, Dickinson, \& Persson 
1994). We found the surface density to be the same within the errors in all 5 
fields studied, and we reported a surface density of ``robust'' candidates for 
Lyman break galaxies of $\approx 0.5$~arcmin$^{-2}$ to an apparent magnitude 
limit of ${\cal R}=25$. 

Since the completion of Paper III, we have been working on extending 
the $U_nG{\cal R}$ imaging technique to more random (i.e., non--QSO) fields and,
most importantly, on following up the candidates with deep spectroscopy 
on the Keck telescope. In this paper we report the successful
results of our first attempts at spectroscopy of the Lyman break candidates, 
new $U_nG{\cal R}$ imaging in (much larger) random fields which confirms
the surface density estimates given in Paper III, 
and the first measurements of $K$--band magnitudes for the $z>3$ population. 

\section{SPECTROSCOPY}

The spectra were obtained during the 
nights of 1-2 October 1995 (UT) and 29-30 October 1995 (UT) 
with the W.M. Keck Telescope and 
the Low Resolution Imaging Spectrograph (Oke \et 1995). 
By using multi-object slit masks, 
each $\sim 4$\arcm\ by 7\arcm\ (and in one case long--slit spectroscopy), 
we recorded the spectra of 25
robust candidates for $z >3$ galaxies
in three fields: two are QSO fields,  0000$-$263 and 0347$-$383, and the third 
is the Hawaii deep survey field SSA~22 (Lilly \et 1991; Cowie \et 1994). 
These observations were made in far from optimum 
conditions, being either at large airmass (typically 
1.8-2.0 for the southern fields) or in grey time (SSA~22); consequently, the 
spectra we obtained do not reflect the full capabilities of the LRIS+Keck 
combination.

Candidates in the 0000$-$263 and 0347$-$383 fields were selected on the basis of the
photometry in Papers I and III, augmented by deeper images of the former field acquired
more recently. For the 
SSA~22 field we used photometry 
obtained with the Palomar 5m telescope
in 1995 August and described in \S 4 below.
As the number of available slitlets exceeded the number of 
robust candidates which could be observed with each mask,
we also observed four ``marginal'' candidates---that is objects whose
colors placed them at or near the boundaries of the color criteria 
(in either $G-{\cal R}$ or $U_n-G$) outlined in Paper III  
---for the purpose of possibly 
improving our selection criteria in future. 
We summarize the results of the spectroscopy in Table 1, which includes relevant 
data for all of the candidates, both robust and marginal, which have been
attempted to date.   
 
Of the 23 objects identified, 16 are galaxies with $3.01 \le z \le 3.43$ and
3 are faint QSOs. 
The other 4 identified objects were considered to be marginal candidates,
in 3 cases because of red $G-{\cal R}$ colors and significant $U_n$ band
detections (all three turn out to be halo subdwarf stars). The 4th case,
which showed evidence for a spectral break but which had a $U_n-G$ color
slightly bluer than our robust candidates, is a galaxy at $z=2.780$. 
Of the 8 robust candidates with undetermined redshifts, none is
actually {\it inconsistent} with being at $z > 3$; in no case do we find
a secure spectral feature at a lower redshift (the spectral range covered
includes [OII]$\lambda 3727$ up to $z<1.4$).
Many of these objects exhibit continuum
discontinuities ($\sim 0.3-0.6$ magnitudes)
which are indicative of the onset of the Lyman
$\alpha$ forest. 

To summarize, roughly 70\% of the objects which we considered
to be robust candidates have been confirmed to have
$3.0 \le z \le 3.5$, and the other $\sim$30\% are consistent with
the same range of redshifts.\footnote{We note that an additional
Lyman break candidate from our survey has recently been confirmed 
spectroscopically.
In Paper III we proposed that an ${\cal R}=25.0$ galaxy 2\secpoint9
from the line of sight to Q2233+1310 is the $z=3.15$ damped Lyman $\alpha$
absorber, as this is the best candidate Lyman break object in the small
field we observed around the QSO. Djorgovski \et (1996) have 
serendipitously
obtained a spectrum of this candidate showing
Lyman $\alpha$ emission at the predicted	
wavelength.}

\section {THE SPECTRA OF $z>3$ GALAXIES}

In Figure 1 we have reproduced some examples of Keck spectra of $z > 3$ 
galaxies, chosen to illustrate the variety of features encountered.
In each case we have included for comparison a recent $HST$ spectrum of 
the central starburst region in the Wolf-Rayet galaxy NGC~4214
(Leitherer \et 1996).

Qualitatively, 
the similarity between the high-redshift galaxies and local examples 
of starbursts is striking. In each case the dominant
characteristics of the far-UV spectrum are: 
{\it (i)} a flat continuum; 
{\it (ii)} weak or absent Lyman $\alpha$ emission; 
{\it (iii)} prominent high-ionization stellar lines of He~II, C~IV, Si~IV and N~V
and {\it (iv)} strong interstellar absorption lines due to low-ionization 
stages of C, O, Si and Al.
These stellar and interstellar 
lines are the most distinctive spectral features in many of the $z >3$ galaxies;
this makes confirmation of objects at $z > 2$ with optical spectroscopy
much more promising than might have been anticipated, even in the absence 
of strong ultraviolet emission lines.
The continuum specific luminosity at 1500 \AA\ of a typical
$z >3$ galaxy in our sample is 
$L_{1500} = 10^{41} h_{50}^{-2}$~erg~s$^{-1}$~\AA$^{-1}$
(for $q_0 = 0.5$; 3 times greater for $q_0 = 0.05$);
this is $\approx 500 - 1500$ times higher (depending on the value of $q_0$) 
than the knot of star formation in NGC~4214 observed by 
Leitherer \et (1996) and  $\approx 30 - 100$ times higher than the 
brightest such knots seen in nearby starburst galaxies.
Thus, for a `normal' IMF, the
far--UV continuum we see in the  $z > 3$ objects is produced by the 
equivalent of $\approx 2-6 \times 10^5$~O7 stars. 

A detailed discussion of the spectra is beyond the scope of
this paper; here we limit ourselves to a few preliminary considerations.
All of the galaxies in the sample are consistent with unreddened models
of young star forming galaxies (Bruzual \& Charlot 1994).  
The expected dust--free $G-{\cal R}$ colors are in the range $0.3-0.9$ after accounting
for the blanketing of the $G$ band by the Lyman alpha forest (see Madau 1995, Paper III), the same
as the observed range in our sample. 
Modulo the uncertainties in models of young galaxies (see, e.g., Charlot 1996), we
estimate that the {\it maximum} reddening allowed 
in the $G - {\cal R}$ color is $\sim 0.4$ magnitudes. Taking this as an upper limit,
and using the ``extragalactic'' reddening curve of 
Calzetti \et (1994), 
the extinction at observed ${\cal R}$ (rest $\sim 1600$ \AA) 
would be $< 1.7$ magnitudes, or a factor of $< 5$. The corresponding
optical reddening in the rest frame of the galaxies would be $E(B-V) < 0.3$ magnitudes.   

Despite the apparent lack of a large amount of dust, Lyman $\alpha$ emission is always 
{\it much} weaker (usually by factors of more than 10)
than the ionization--bound,
dust--free expectations, given the production of ionizing photons by
massive stars which we measure directly from the UV continuum.
Reasons for the preferential extinction of Lyman $\alpha$ emission have been discussed by, e.g., 
Charlot \& Fall (1993) and Chen \& Neufeld (1994); our observations are
entirely consistent with the same low (but non--zero) dust content
inferred to be present in the high redshift damped Lyman $\alpha$ aborption systems
(e.g., Pei \et 1991, Pettini \et 1994).  
In the spectra which do show Lyman $\alpha$ emission the typical 
rest--frame equivalent width is 
$W_{{\rm Ly}\alpha} = 3-20$ \AA, but apparently
most $z>3$ galaxies have Lyman $\alpha$ emission lines weaker than these 
values, in striking similarity to nearby star--forming galaxies (e.g., Giavalisco,
Koratkar, \& Calzetti 1996). 
For example, C23 0000-263 in Figure 1 
has among the largest implied star formation
rates in our sample (as discussed in \S 6 below), and yet has no measurable
Lyman $\alpha$ emission.

The Lyman $\alpha$ emission lines and number density that we measure for our 
galaxies are consistent with the essentially null results of all previous 
searches for high-$z$ galaxies based on this spectral feature (e.g., Thompson, 
Djorgovski, \& Trauger 1995; Lowenthal \et 1995; Pettini \et 1995). 
Given the star formation rates we derive, the space density of the Lyman
break galaxies (\S 6) is also consistent with null results from near--IR 
surveys (e.g., Pahre \& Djorgovski 1995; Mannucci \et 1994) 
for redshifted H$\alpha$, [OIII], and [OII] lines. 
While the
near--IR surveys can reach star formation rates comparable to those 
we observe, the volumes surveyed to date are too small for a significant
chance of detection. 

While there is a great deal of variety in the
strengths of the lines which are predominantly of stellar origin (C~IV
$\lambda 1549$, Si~IV $\lambda\lambda 1393$,1402 and
He~II $\lambda 1640$), we find these features to be generally weaker
than in the spectra of present-day starbursts. This is probably an 
abundance effect; these lines are formed predominantly in the winds of 
massive stars, where both mass-loss rates and wind terminal velocities are 
known to depend sensitively on metallicity (e.g. Walborn \et 1995). 

The strongest interstellar lines indicated in Figure 1 have typical 
rest-frame equivalent widths $W_0 \simeq 2-3.5$~\AA. While the interstellar 
medium of these galaxies has obviously undergone some chemical enrichment, 
it is not possible to deduce metallicities from our spectra, since the 
absorption lines in question are undoubtedly heavily saturated. Under these 
circumstances, values of metallicity anywhere
between 1/1000 of solar and solar are compatible with the line strengths and 
higher-resolution observations of intrinsically weaker lines are 
required to measure element abundances (Pettini \& Lipman 1995).
The equivalent widths of saturated absorption lines are much more 
sensitive to the velocity dispersion of the gas than to its column density.
The observed $W_0 \simeq 2-3.5$~\AA\ correspond to FWHM $ \geq 400 - 
700$~\kms\ which in turn imply approximate velocity dispersions 
$\sigma = 180 - 320$~\kms 
(slightly higher values apply if rotation is the dominant effect). In any case, 
if these velocity spreads reflect primarily 
gravitationally induced motions in the large-scale interstellar medium of 
the galaxies, the masses implied are comparable to those of present-day 
luminous galaxies. Smaller masses would result if interstellar shocks,
local to the star-forming regions, contribute to the line widths. 
Spectra of higher S/N and resolution are required to resolve 
this question.

\section {THE SURFACE DENSITY OF LYMAN BREAK OBJECTS}

In parallel with this spectroscopic follow-up, we have been obtaining 
deep $U_nG{\cal R}$ images of additional fields using the COSMIC camera 
at the prime focus of the Palomar 5~m telescope with two principal aims:
{\it (i)} Improve the statistics on the surface density of Lyman break 
objects, by taking advantage of the much larger field of view available
to COSMIC (9\minpoint7 $\times$ 9\minpoint7, compared to approximately 
1\minpoint5 $\times$ 1\minpoint5 of most of the observations reported in 
Paper~III); and {\it (ii)} Extend this work to random fields---that is 
fields which do not include known QSOs---so as to assess if the surface 
density deduced in Paper~III is typical of the general field. 

To this end we have chosen one of the Hawaii deep survey fields, 
SSA 22, which has also been observed extensively with 
post-refurbishment $HST$ (e.g., Cowie, Hu, \& Songaila 1995a)
in several pointings, all falling within our Palomar 5~m images.
These new $U_nG{\cal R}$ images reach
depths comparable to those of the data in Paper III, and allow us to 
be complete in our identification of 
$z >3$ galaxies down to ${\cal R} =25.0$. We find a total
of 31 robust candidates satisfying the selection criteria 
outlined in Paper III out of a total of 2340 objects  
detected to the same apparent magnitude level (the field size in the
stacked images was 8\minpoint8 $\times$ 9\minpoint0).
We therefore deduce a surface density of Lyman break
candidates of $0.40 \pm 0.07$~arcmin$^{-2}$ 
(or $1.44 \times 10^3$ per square degree), consistent with 
(and significantly more accurate than) the value
$\approx 0.5$~arcmin$^{-2}$ reported in Paper III. 

We conclude that: {\it (i)} 
Lyman break objects in the redshift range $3.0 \le z \le 3.5$
represent about 1.3\% of all objects to ${\cal R}=25.0$, and
2.0\% of all objects in the magnitude range $23.5 \le {\cal R} \le 25.0$; 
and {\it (ii)} the density of $z > 3$ galaxies is 
not significantly higher in fields including bright high-$z$ QSOs.

\section {THE NEAR--IR EMISSION FROM $z>3$ GALAXIES}

In 1995 October we obtained deep $K_s$ band images of a small
subset of the $z>3$ candidates 
using the Near Infrared Camera (Mathews \& Soifer 1994) on the W. M. Keck telescope. 
A total of 5 candidates was observed, with typical integration times of
6000s; 4 of the candidates are among those confirmed spectroscopically (see Table 1).
The measured $K$ magnitudes (typically measuring rest--frame B or V of the galaxies)
of the non--AGN candidates (0000$-$263 C09
appears to be an AGN) range from $K=21.3-22.1$, with
$2.3 \le {\cal R} -K \le 3.2$\footnote{The optical passbands are on an
``AB''--normalized system, whereas the $K$ magnitudes are the standard
system. A flat spectrum ($f_{\nu}=$const.) source
has ${\cal R}-K=1.85$} . These colors are consistent with models of continuous
star formation which could have begun as early as $\simgt 1$ Gyr prior to the epoch
at which we observe them and are redder than would be expected if we were
seeing instantaneous ``bursts'' of star formation (Bruzual \& Charlot 1993).    
A consequence of adopting the {\it maximum} amount of reddening allowed by the
UV colors of the galaxies (\S 4) is a reddening in ${\cal R} - K$ of
$\sim 0.8$ magnitudes; this would lower the ages significantly and
formally allow single burst models
younger than a few times 
$10^7$ years. We regard such short
lifetimes as unlikely as they would imply that we are seeing large numbers
of galaxies all bursting simultaneously. 

\section {IMPLICATIONS}

The confirmation of a substantial population of luminous, star--forming
galaxies at $z > 3$ has considerable implications for our understanding
of galaxy formation and evolution. One of the advantages of searching for 
high$-z$ galaxies in the optical is that one observes directly
the far--UV continuum produced by early-type stars 
(our ${\cal R}$ bandpass samples the rest frame continuum
at $\lambda_0 \sim 1600$ \AA). 
Consequently, in the absence of dust (see \S 5), relatively minor assumptions 
are necessary to deduce
the formation rate of massive stars and the accompanying production
of ionizing photons; this is {\it not} the case for
measurements of [OII] line luminosities
which are subject to uncertainties as large as a factor of $\sim 5$ 
(see Gallagher, Bushouse, \& Hunter 1989). 
For the purpose of
estimating star formation rates from the observed UV luminosities,
we have made use of the calculations by Leitherer, Robert, \& Heckman 
(1995) which are based on ultraviolet libraries of massive star spectra 
coupled with an evolutionary synthesis code. We have used the 
``continuous star formation'', rather than ``single burst'', models; we 
consider the former case more likely, for reasons given in \S 5. 
For a Salpeter initial mass function with an upper mass cut-off of 80 
M$_{\sun}$,\footnote{
The parameters of the IMF
are relevant for extrapolating the observed rate of formation of 
massive and short-lived stars to the {\it total} SFR;
for our purposes the slope of the IMF is the important parameter, while
the upper mass cut-off has only a minor effect---see Table 2 of Leitherer 
\et 1995. } 
a SFR =  $ 1M_{\sun}$~yr$^{-1}$ produces a
luminosity at 1500 \AA\ (in the rest-frame) 
$L_{1500} = 10^{40.1}$~erg~s$^{-1}$~\AA$^{-1}$. 
At $z = 3.25$ (in the middle of the redshift 
range to which we are sensitive to 
the detection of Lyman break galaxies), 
an apparent ${\cal R} = 24.5$ (on the AB system)
corresponds to $L_{1500} = 10^{41.1}$~erg~s$^{-1}$~\AA$^{-1}$ 
($H_0 = 50$~km~s$^{-1}$~Mpc$^{-1}$) for $q_0 = 0.5$, and 
$L_{1500} = 10^{41.6}$~erg~s$^{-1}$~\AA$^{-1}$ for $q_0 = 0.05$\,.  
The population of Lyman break galaxies we detect is in the range
$25.5 \simgt {\cal R} \simgt 23.5$;
therefore, the implied star formation
rates range from $(4-25)h_{50}^{-2}$ M$_{\sun}$ yr$^{-1}$ ($q_0=0.5$) to
$(12-75)h_{50}^{-2}$ M$_{\sun}$ yr$^{-1}$ ($q_0=0.05$), with the weighted
average being $8.5h_{50}^{-2}$ ($25h_{50}^{-2}$) M$_{\sun}$ yr$^{-1}$ 
for $q_0=0.5$ (0.05).

Assuming that we have uniformly probed the redshift
range $3.0 \le z \le 3.5$ in our surveys (an assumption that is supported by
the results of the spectroscopy), the co-moving density of the star--forming
galaxies is at least 
$3.6 \times 10^{-4}h_{50}^{3}$ ($6.7 \times 10^{-5}h_{50}^{3}$) 
Mpc$^{-3}$ for $q_0=0.5$ (0.05), or about 1/2 (1/10) of 
the space density of present--day
galaxies with $L > L^{\ast}$ for $q_0=0.5 (0.05)$ 
(Ellis \et 1996). 
The total star formation rate
per co--moving volume produced by the {\it observed} population 
at $z>3$ is then
$3.1 \times 10^{-3}h_{50}$ ($1.8\times 10^{-3}h_{50}$) 
M$_{\sun}$ yr$^{-1}$ Mpc$^{-3}$
for $q_0=0.5$ (0.05).
These numbers must be viewed as strict lower limits on the {\it total} star formation 
rate at $z > 3$---the fraction arising in only the most actively star--forming objects;
taken at face value, the star formation density that we observe is only
$\sim 25\%$ of the total star formation rate seen at the {\it present} epoch
(Gallego \et 1995).
For comparison, the star formation rates (per object) and star formation 
densities recently reported by 
Cowie, Hu, \& Songaila (1995b) at $z \simgt 1$
are approximately the same as we have observed at $z \simgt 3$; however, as
we detail in Giavalisco, Steidel, \& Macchetto (1996), the morphology of the objects hosting the
large star formation rates seems to be entirely different in the two redshift regimes---
the great majority of the $z>3$ objects have very compact morphologies, with
half--light radii in HST images of $0.2-0.4$\arcs ($1.5-3h_{50}^{-1}$ kpc for
$q_0=0.5$), and in most cases are relatively  
``regular'' in appearance. 
As pointed out in Paper III, it will be very interesting to observe how
rapidly the surface density of Lyman break objects increases at fainter apparent
magnitudes. 

In summary, we have discovered a population of star--forming galaxies
with redshifts $3.0 \le z \le 3.5$ using a color--selection technique whose
efficiency is very high, allowing the first {\it systematic} study
of the nature of galaxies at such large redshifts.  
The space density, star formation rates, morphologies and physical 
sizes (see Giavalisco \et 1996), 
masses, and early
epoch of the galaxy population that we have isolated all
support the possiblity that we are seeing directly, for the first time, 
the formation of 
the spheroidal component in the progenitors of 
present-day luminous galaxies.
The fact that we detect such a substantial population 
using a flat--UV spectrum selection criterion suggests that dust obscuration
may {\it not} be 
an important limiting factor 
in searches for high-redshift galaxies. 
In any case, our results 
demonstrate beyond any doubt that 
massive galaxy formation was well underway
by $z=3.5$.  

\acknowledgements

We would like to thank the staff of the W.M. Keck Observatory, without whom
these observations would not have been possible. CCS acknowledges support
from the Sloan Foundation and from the NSF through grant AST--9457446. We thank
W. Sargent, M. Fall, and A. Phillips for comments on an earlier 
draft of the manuscript. We are especially
grateful to Claus Leitherer for providing us with the spectra of nearby
star--forming galaxies prior to publication.

\begin{deluxetable}{lccrccr}
\tablewidth{0pc}
\scriptsize
\tablecaption{Summary of Candidates Observed Spectroscopically}
\tablehead{
\colhead{Object} & \colhead{${\cal R}$} & \colhead{$G-{\cal R}$} & \colhead{$U_n-G$} &
\colhead{${\cal R}-K$} &
\colhead{Redshift\tablenotemark{a}} & \colhead{$W_{\lambda}$(Ly $\alpha$)\tablenotemark{b}} } 
\startdata
0347$-$383 N5   &  23.82 & 0.65 & $>$2.85 &  2.5  & 3.243 & 6 \nl
0000$-$263 C02\tablenotemark{M}  &  24.06 & 1.23 & $>$2.45 & \nodata &  star & \nodata \nl        
0000$-$263 C04  &  23.71 & 1.00 & $>$2.96 & \nodata & 3.789 & (QSO) \nl 
0000$-$263 C06  &  25.49 & 0.96 & $>$1.75 & \nodata & 3.202   & 3  \nl 
0000$-$263 C09  &  24.44 & 0.00 & $>$3.36 &  3.9     & 3.428\tablenotemark{c} & 355 \nl
0000$-$263 C11  &  25.30 & 0.50 & $>$2.32 &  3.2     & 3.135   & 8   \nl 
0000$-$263 C13  &  25.38 & 0.90 & $>$1.75 & \nodata & 3.238   & $<$0 \nl
0000$-$263 C14  &  24.54 & 0.77 & $>$2.45 & \nodata & 3.022   & 7  \nl
0000$-$263 C16  &  24.47 & 0.85 & $>$2.54 & \nodata & 3.056   & 18 \nl
0000$-$263 C17  &  25.35 & 0.68 & $>$1.97 & \nodata & 3.169   & $<$0  \nl
0000$-$263 C19  &  25.26 & 0.99 & $>$2.17 & \nodata & \nodata  & \nodata \nl 
0000$-$263 C21\tablenotemark{M} &  23.79 & 1.24 &    1.85 & \nodata & star  & \nodata \nl
0000$-$263 C23  &  23.93 & 0.51 &    1.73 & \nodata & 3.199 & $<$0 \nl
0000$-$263 C25  &  25.08 & 0.60 & $>$1.78 & \nodata & 3.342 & $<$0 \nl
0000$-$263 C26  &  25.77 & 0.26 & $>$2.13 & \nodata & \nodata & \nodata   \nl
0000$-$263 C27\tablenotemark{M} &  25.00 & 0.52 &    1.52 & \nodata & 2.780   & $<$0 \nl
0000$-$263 C28  &  24.90 & 0.70 & $>$1.90 & \nodata & \nodata & \nodata \nl
SSA22 C01       &  25.50 & 0.25 & $>$1.93 & \nodata & \nodata & \nodata \nl
SSA22 C03       &  25.37 & 0.12 & $>$1.95 & \nodata & \nodata & \nodata \nl
SSA22 C10       &  24.86 & 0.21 & $>$2.26 & \nodata & 3.375    & $<$0 \nl
SSA22 C11       &  24.82 & 0.41 & $>$2.18 & \nodata & \nodata & \nodata\nl
SSA22 C12       &  24.78 & 0.43 & $>$2.13 & \nodata & 3.201  & 3 \nl
SSA22 C14       &  24.64 & 0.26 & $>$2.31 & \nodata & 3.401  & $<$0 \nl
SSA22 C16       &  24.62 & 0.46 & $>$2.10 & \nodata & \nodata & \nodata \nl
SSA22 C19       &  24.58 & 0.39 & $>$2.50 & \nodata & 3.019  & 22 \nl
SSA22 C20       &  24.52 & 0.27 & $>$2.55 & \nodata & 3.019  & $<$0\nl
SSA22 C24       &  23.45 & 0.38 & $>$3.01 & \nodata & \nodata & \nodata \nl
SSA22 D10       &  21.75 & 0.52 &    2.56 & \nodata & 3.083 & (QSO) \nl
SSA22 D11\tablenotemark{M}  &  21.53 & 1.13 &    3.26 & \nodata & star  & \nodata\nl
SSA22 D12       &  20.98 & 0.87 &    3.82 & \nodata & 3.352  & (QSO) \nl
2233+131 N1     &  25.02 & 0.15 & $>$2.38 & [3.0]    & 3.151\tablenotemark{d}  & \nodata \nl
\enddata
\tablenotetext{a}{Spectroscopic redshift, when secure. }
\tablenotetext{b}{Rest equivalent width of Lyman $\alpha$ emission line for
galaxies, in \AA; values
$<$0 indicate net absorption at the position of Lyman $\alpha$}
\tablenotetext{c}{ Object ``G2'' (Giavalisco \et 1994)-- Keck spectrum shows presence
of C~IV, He~II, O~VI, and N~V emission, hence G2 is probably an AGN}
\tablenotetext{d}{Spectroscopic confirmation from Djorgovski \et (1996)}
\tablenotetext{M}{Marginal candidate on the basis of $Un-G$ and $G-{\cal R}$ colors}
\end{deluxetable}
\newpage 
\begin{figure}
\figurenum{1a,b}
\epsscale{0.9}
\plotone{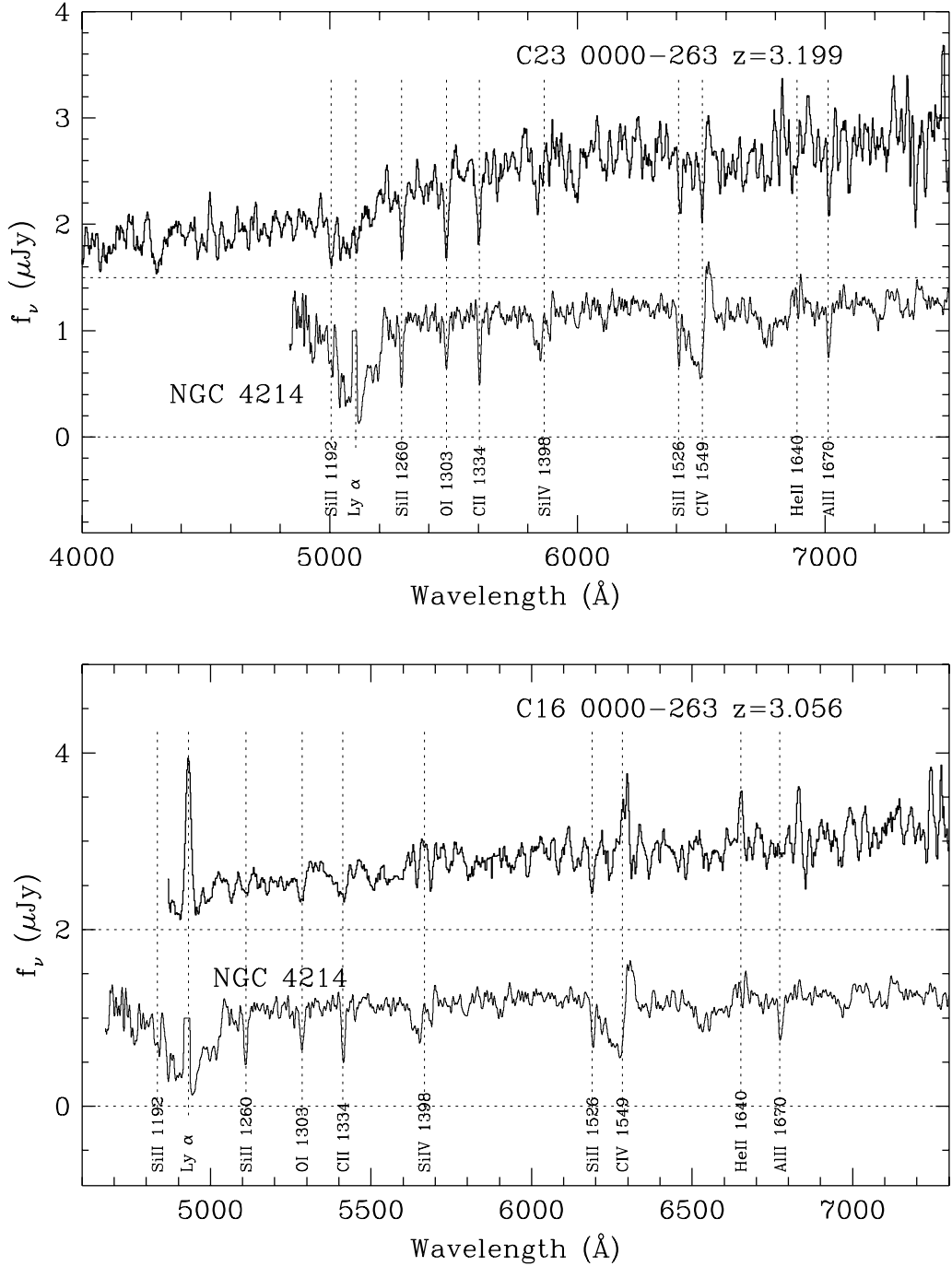}
\caption{Examples of Keck spectra
of $z > 3$ galaxies. Below each observed spectrum, we have plotted the spectrum
of a star forming ``knot'' in NGC 4214, a nearby starburst galaxy (Leitherer \et
1996), 
after shifting the spectrum
to the measured redshift of the high-$z$ galaxy. Some of the
spectral features seen in local star--forming regions (both stellar and
interstellar) have been labeled, for comparison with
the high redshift objects 
(see text for
discussion).} 
\end{figure}
\newpage
\begin{figure}
\epsscale{0.9}
\figurenum{1c,d }
\plotone{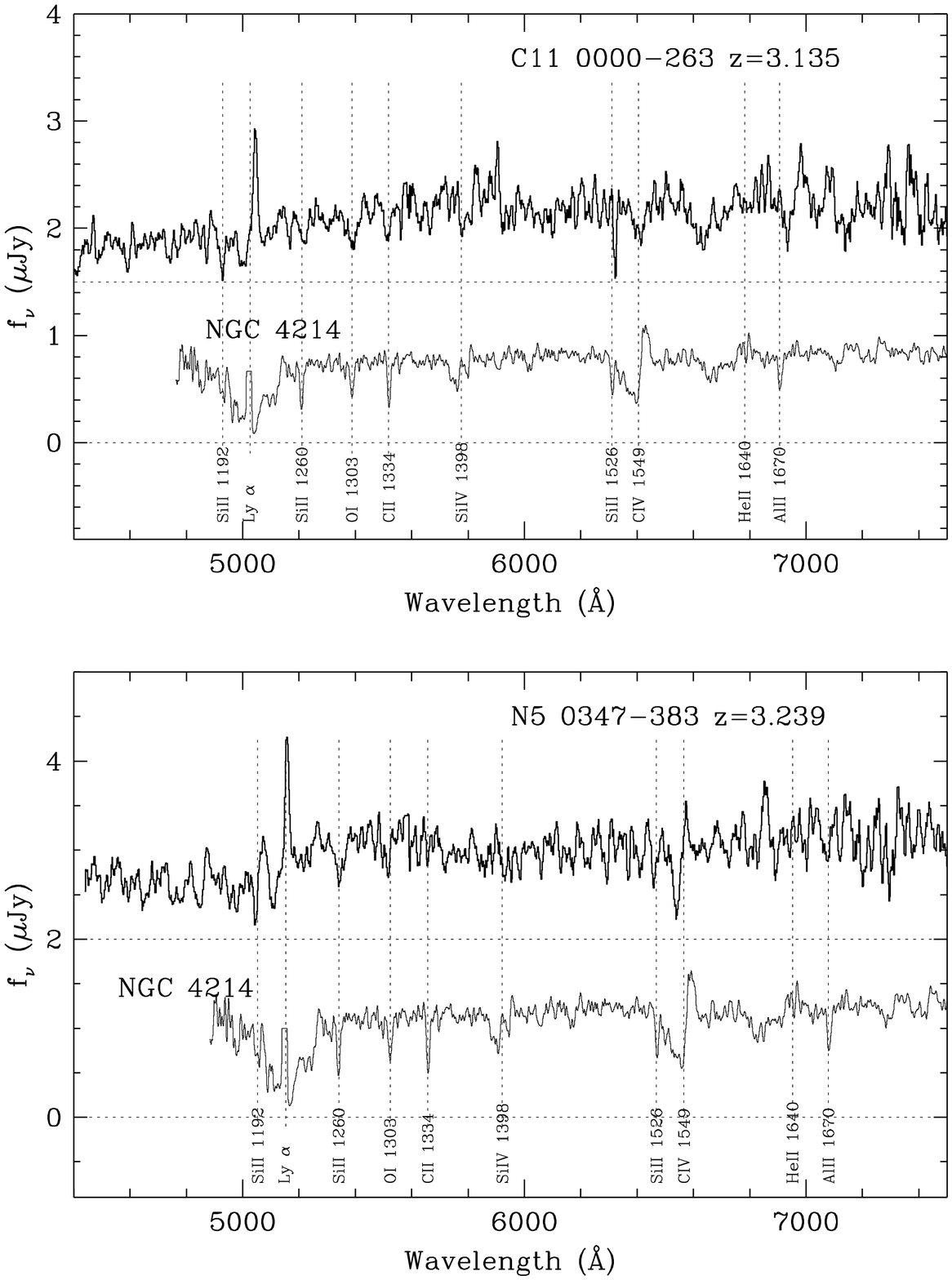}
\caption{(continued)}
\end{figure}

\end{document}